# Revealing unforeseen diagnostic image features with deep learning by detecting cardiovascular diseases from apical four-chamber ultrasounds

Li-Hsin Cheng[1,+], MSc, Pablo B.J. Bosch[2,+], MSc, Rutger F.H. Hofman[2], PhD, Timo B. Brakenhoff[3], PhD, Eline F. Bruggemans[4], MSc, Rob J. van der Geest[1], PhD, Eduard R. Holman[5,*], MD PhD

[1] Division of Image Processing, Department of Radiology, Leiden University Medical Center, Leiden, the Netherlands

[2] Department of Science, Vrije Universiteit Amsterdam, Amsterdam, the Netherlands

[3] Ynformed, Utrecht, the Netherlands

[4] Department of Cardiothoracic Surgery, Leiden University Medical Center, Leiden, the Netherlands

[5] Department of Cardiology, Leiden University Medical Center, Leiden, the Netherlands

[+] These authors contributed equally to the work.

* Corresponding author:

   Eduard R. Holman, MD PhD

   Department of Cardiology

   Leiden University Medical Center

   Albinusdreef 2

   PO Box 9600

   2300 RC Leiden, the Netherlands

   E-mail: E.R.Holman@lumc.nl

# Abstract

**Background**. With the rise of highly portable, wireless, and low-cost ultrasound devices and automatic ultrasound acquisition techniques, an automated interpretation method requiring only a limited set of views as input could make preliminary cardiovascular disease diagnoses more accessible. In this study, we developed a deep learning (DL) method for automated detection of impaired left ventricular (LV) function and aortic valve (AV) regurgitation from apical four-chamber (A4C) ultrasound cineloops and investigated which anatomical structures or temporal frames provided the most relevant information for the DL model to enable disease classification.

**Methods and Results**. A4C ultrasounds were extracted from 3,554 echocardiograms of patients with either impaired LV function (n=928), AV regurgitation (n=738), or no significant abnormalities (n=1,888). Two convolutional neural networks (CNNs) were trained separately to classify the respective disease cases against normal cases. The overall classification accuracy of the impaired LV function detection model was 86%, and that of the AV regurgitation detection model was 83%. Feature importance analyses demonstrated that the LV myocardium and mitral valve were important for detecting impaired LV function, while the tip of the mitral valve anterior leaflet, during opening, was considered important for detecting AV regurgitation.

**Conclusion**. The proposed method demonstrated the feasibility of a 3D CNN approach in detection of impaired LV function and AV regurgitation using A4C ultrasound cineloops. The current research shows that DL methods can exploit large training data to detect diseases in a different way than conventionally agreed upon methods, and potentially reveal unforeseen diagnostic image features.

**Keywords**: explainable artificial intelligence, echocardiography, apical four-chamber cineloop, 3D convolutional neural network, impaired left ventricular function, aortic valve regurgitation

# Introduction

Echocardiography is the main diagnostic imaging modality for the assessment of cardiovascular disease (CVD). However, although it is applicable in most settings, interpretation of echocardiograms is time consuming and subject to intra- and inter-observer variability. In addition, the image interpretation requires experienced experts, which are not always accessible. With the prevalence of CVD increasing, a scarcity of expert cardiologists to perform high quality assessments is expected (1). With the rise of highly portable, wireless, and low-cost ultrasound devices and automatic ultrasound acquisition techniques (2), the availability of an automated interpretation method requiring only a limited set of views as input could make echocardiography based CVD diagnosis more accessible. Such a system could become beneficial in geographic regions with limited access to expert cardiologists and sonographers. It could also support general practitioners in the management of patients with suspected CVD, facilitating timely diagnosis and treatment of patients.

Recent developments in artificial intelligence (AI) technology provide an opportunity to achieve this goal. In particular, deep learning can automatically learn a hierarchy of features from a huge amount of image data (3), thus having the potential to uncover diagnostic features in the data not previously recognized. Successful deep learning algorithms have already been developed to facilitate various steps in the workflow of echocardiography interpretation (4–6). Among the models developed, 3D convolutional neural networks (CNNs) do allow the analysis of both spatial and temporal information of the input (7–9). Therefore, in this study, we developed models to detect cardiovascular diseases with a 3D CNN based approach taking cineloops as input.

As a pilot study for a general-purpose automated CVD diagnosis model using simple inputs, we adopted the apical four chamber (A4C)-view ultrasound cineloop as the input data, as we consider A4C a general view that contains comprehensive information in a single shot. We chose two abnormalities for the models to learn, namely detection of impaired left ventricular (LV) function against normal, and detection of aortic valve (AV) regurgitation against normal. Impaired LV function can be seen on the A4C view, but is usually determined together with other apical viewpoints and 2 to 3 parasternal viewpoints (10,11). This task would allow us to verify the feasibility of the 3D CNN approach, distinguishing the abnormality with limited but highly relevant information. On the other hand, AV regurgitation is typically diagnosed based on color Doppler images using one or more viewpoints (12). This detection task would allow us to further test the limit of a 3D CNN in distinguishing an abnormality that is not obvious on the A4C view. At the same time, it allows investigating whether the model identifies unforeseen image features while detecting the abnormality with an approach different from the clinical convention. Therefore, after training the models, we performed feature importance analysis to try to inspect what are the identified image features associated with each diagnostic task (Figure 1).

The current study proposes the use of deep learning to automatically derive CVD diagnoses from echocardiography cineloops with two specific focuses. Firstly, we aimed to investigate the feasibility of using 3D CNNs to detect diseases based solely on the A4C view. Secondly, through feature importance analysis, we aimed to investigate whether the built models can identify anatomical and motion related image features associated with the diseases, typically not being considered in conventional image interpretation.

## Methods

### Data Extraction

Echocardiographic data appropriate for this retrospective study was anonymously extracted from the echocardiography database of the Heart Lung Center at the Leiden University Medical Center (LUMC), Leiden, the Netherlands. The study was approved with waiver of informed consent by the Ethics Committee of the institution.

All patients underwent echocardiography in the left lateral decubitus position, using a commercially available system (Vivid 7, E9 or E95; GE Vingmed Ultrasound AS, Horten, Norway) and 3.5 MHz transducers. Standard M-mode and 2-dimensional, color, pulsed, and continuous wave Doppler images were acquired according to the recommendations of the European Association of Echocardiography (13). Offline analysis was performed using EchoPAC (version 203.59.0; GE Medical Systems). Only echocardiographic data of patients that were diagnosed as *normal, impaired LV function,* or *AV regurgitation* were included in this study. For patients diagnosed as *normal*, the images showed no significant abnormalities in anatomy or motion. Patients with *impaired LV function* showed a reduced LV wall motion, either attributable to an ischemic or non-ischemic cause. The impairment was staged by human experts as either mild or severe, which were the two impaired LV function classes defined for our classification task. *AV regurgitation* was diagnosed based on the backflow of the blood from the aorta to the left ventricle in a Doppler ultrasound, using color flow mapping and spectral Doppler. The regurgitation was staged by human experts as either mild, moderate, or severe. For the classification task, we defined two classes: the mild class and the combined class of moderate and severe regurgitation (from here on called the "substantial" AV regurgitation class).

Examinations for each category were anonymously extracted in DICOM-format. As such, meta data like sex, age, and weight of patients were unknown. From the available database, a dataset was created by manually selecting all A4C cineloops. The resulting 3,554 ultrasounds were randomly split into separate datasets for training (70%), validation (10%), and testing (20%). Unfortunately, due to anonymization of the data, we were not able to include the ultrasounds of individual patients in a single dataset and enforce sample independence for patients. Table 1 shows per class the total number of extracted ultrasounds and ultrasounds per dataset for training, validation, and testing.

**Data Preprocessing and Augmentation**

In each ultrasound image acquisition, the frame rate and heart rate are different and thus the number of frames for a single cardiac cycle. To make the temporal dimension of each input video to the model consistent, all ultrasounds were resampled such that the duration of one cardiac cycle was captured in 30 frames. During training time, we sampled the clip with a random starting point on-the-fly. It is a part of the data augmentation to increase data diversity, analogous to random cropping in the spatial dimension. During validation and test time, the starting point of the clip was always set as the starting point pre-determined by the software of the ultrasound scanner.

The raw ultrasound is embedded with ECG and text annotations. We filtered out all the embedded information so that our models would learn the two diagnoses based solely on the actual image information. The intensity of the filtered ultrasounds were subsequently aligned to the global intensity distribution of the whole dataset through histogram matching (14). This procedure helps ensure that the brightness and contrast of each video were roughly the same.

From the raw ultrasound of 708 × 1,016 pixels, we cropped the center 549 × 549 pixels containing the fan-shaped field of view, then down-sampled the image to 112 × 112 pixels. During training, random translation and rotation were applied on-the-fly as augmentations to increase data diversity, which mimic variations that would happen in real world due to different angles and positions of the transducer.

**Model Development**

We built two 3D CNNs to separately classify impaired LV function and AV regurgitation cases against the normal cases. We decided to adopt the R(2+1)D model architecture (15) for the tasks. This particular architecture decomposes a 3D convolution into a spatial convolution followed by a temporal convolution, which was recently used to successfully predict the ejection fraction based also on A4C ultrasound cineloops (7).

We used cross entropy as the loss function and the ADAM optimizer to update network weights. The learning rate was set at 0.001 and the batch size was 16. Early stop with a patience of 50 epochs was applied. The models were implemented with the deep learning library Pytorch 1.7, and the training was performed on a NVIDIA Quadro RTX 6000 GPU. The code and the trained models are made available on GitHub: https://github.com/LishinC/Disease-Detection-and-Diagnostic-Image-Feature.

**Feature Exploration using tSNE Visualization**

T-Distributed Stochastic Neighbor Embedding (tSNE) (16) is a dimensionality reduction method that is often used to embed high dimensional data into a two (or three) dimensional embedding for the purpose of visual exploration. When performing embedding, the tSNE method tries to preserve the local relative distance between samples, such that closer data points in a tSNE plot imply more similar samples.

tSNE can be applied to visualize a variety of high dimensional data. In this study, the method was used to visualize both the input video and the features extracted by the trained 3D CNN (512-dimensional vector). tSNE visualization of the input video shows the similarity of samples based directly on the pixel intensities, while visualization of the extracted features shows the similarity of samples based on the feature values. Comparison of the two can reveal whether the 3D CNN successfully extracted features relevant to the diagnosis and filtered out the noise, such that samples belonging to the same diagnosis were clustered closer together in the extracted-feature plot as compared to the input-video plot. Additionally, the extracted-feature plot could also indicate the relationship (the distance/similarity to each other) between several overlapping clusters.

**Feature Importance Analysis**

We utilized the feature importance analysis method DeepLIFT (17,18) to try to decrypt the reasoning behind the models' predictions, which could potentially help reveal diagnostic image features not considered before. DeepLIFT attributes a model's classification output to certain input features (pixels), which allows us to understand which region or frame in an ultrasound is the key that makes the model classify it as a certain diagnosis.

DeepLIFT decomposes the output activation difference between the input and the baseline as a sum of layer-wise relevance values, thus obtaining the contribution of each input feature (pixel) to the output prediction. Upon a given input of interest, DeepLIFT would return an analysis result with the same size as the input for each class. The values in the analysis result reflect the importance of that pixel to the class, and the sign of the values indicates either a positive or negative contribution to the class.

## Results

### Model Performance

Figure 2 summarizes the predictive performance of the two models. Figure 2(A, C) shows the normalized confusion matrices for the impaired LV function and AV regurgitation detection models, respectively. Figure 2(B, D) lists the detailed recall (sensitivity), precision, and F1-score values for each diagnosis for both models.

The impaired LV function detection model achieved an overall accuracy of 86%. The model was able to detect 92% of the ultrasounds classified by the cardiologist as severely impaired function. Of the mildly impaired class, 67% was correctly identified. On the other hand, the AV regurgitation detection model reached an overall accuracy of 83% and was able to detect 71% of the substantial class, but only 25% of the mild class.

### Feature Exploration using tSNE Visualization

Figure 3 shows the tSNE plots for impaired LV function (Figure 3(A, B)) and AV regurgitation (Figure 3(C, D)), with each sample colored by the corresponding diagnosis. By comparing the plots before being processed by the models (Figure 3(A, C)) and after (Figure 3(B, D)), i.e., the input-video plot with the extracted-feature plot, we can observe that samples of the same diagnosis are clustered closer to each other after being processed by the models. This implies that the models have successfully extracted diagnosis-relevant features and filtered out irrelevant noise throughout cascades of convolutional layers.

Especially, for the case of impaired LV function detection, it can be seen in Figure 3(B) that, after processing, the normal, mild, and severe clusters are even ordered in the level of severity. The information about severity, i.e., the correct ordering of the three classes, was actually not given to the model. The one-hot categorical label fed to the model implies only that normal, mild, and severe were three different classes, which does not contain hints about the relative similarity between each class. Therefore, besides being able to differentiate the three classes, the model had also learned the correct relative similarity relationship between the three classes.

**Feature Importance Analysis**

Figure 4 presents the feature importance analysis results produced by DeepLIFT, which attributes (per query) the model's prediction of an output class to certain input features (pixels). For both the impaired LV function and AV regurgitation detection models, we present the analysis calculated based on the same representative normal case. The analysis results are presented as heat maps in Figure 4. The brighter pixels in the heat maps are the input features that positively contributed to the normal class, i.e., image features that made the model classify the case as normal. The highlighted regions can thus be interpreted as the image information that make the models distinguish the normal case from the disease cases. More analyses in video format can be found in our GitHub repository.

For the impaired LV function detection model, DeepLIFT highlighted the basal part of the myocardium from the early systolic to the early diastolic phase. Additionally, the mitral valve was highlighted at the early diastolic phase (Figure 4(A)).

For the AV regurgitation detection model, DeepLIFT highlighted the tip of the mitral valve anterior leaflet, particularly at a short time centering around the moment of valve opening (Figure 4(B)). This indicates that the model not only focused on a specific anatomical structure but also on a specific temporal phase within the cardiac cycle.

# Discussion

As a pilot study to make preliminary diagnoses of cardiovascular diseases automated and thus more accessible, we built two 3D CNNs to detect impaired LV function and AV regurgitation using the A4C-view ultrasound. The impaired LV function model was able to detect 92% of the severe class, and the AV regurgitation model was able to detect 71% of the substantial class. Based on the lower recall, we conclude that detecting AV regurgitation was the more difficult task among the two. This is in line with the fact that AV regurgitation is usually diagnosed using Doppler imaging and not from an A4C view. However, our results also reveal that abnormalities derived from the A4C view, although not obvious to the human eye, were sufficient for the AV regurgitation model to reach an overall detection accuracy of 83%. The success of building the impaired LV function detection model demonstrates the feasibility of deep learning algorithms in identifying the abnormality with limited input information. Furthermore, the success of AV regurgitation detection verifies that the model can detect a disease with an approach different from the current practice, i.e., based upon Doppler information. We attribute this to the models' ability to learn from huge amount of data and derive important features independently.

Using tSNE visualization, we verified that the models had transformed the input ultrasounds into a diagnosis-relevant feature representation. Especially, similar to the reconstruction of disease progression as shown in (19), the impaired LV function detection model had mapped the normal, mild, and severe classes in the correct order. This indicates that the model might have obtained information about the severity of the disease, although to which extent it can accurately rank the severity of each individual case requires further evaluation. Nevertheless, this indicates that the trained model could potentially serve as a tool to systematically evaluate the severity of the disease, which would otherwise be hard to accurately quantify by the human eye, merely from a single view and without additional annotation.

Finally, to see which signs in the input cineloop the models focus on to detect the abnormalities, we analyzed the models with the feature importance analysis method DeepLIFT. The analysis suggests that the mitral valve and the LV myocardium at the basal level are crucial for distinguishing the normal class from impaired LV function. This observation verifies that the model indeed works in a reasonable way to detect disease, since the movement of the myocardium is strongly related to LV function, as is the movement of the mitral valve (20). On the other hand, the analysis suggests that the tip of the mitral valve anterior leaflet, during the opening of the valve, is the most important feature that the model focuses on in order to distinguish the normal class from AV regurgitation. It is possible that the movement of the mitral valve is affected by the abnormal regurgitation jet (12) and hence identified by the model as a key difference. It is also possible that morphological changes, such as mitral valve leaflet enlargement, were the key characteristic that the model used to distinguish cases, as supported by a recent study (21). Although the exact mechanism remains unclear, the analysis shows that certain regions of the heart or phases in the cardiac cycle that people often neglected previously might also have a strong link to the disease. Potentially, a trained model could identify image features that are not yet known to be related to a disease, hence bringing insights to the disease diagnosis.

For the respective models, the DeepLIFT-highlighted regions in the shown normal case represent the general highlighting pattern that we observed in most of the normal cases. These consensus highlighted regions are the diagnostic image features concluded by the respective models after learning from the training data. However, if we input disease cases to the analysis, the highlighted regions in different queries would be quite different. (Examples of the DeepLIFT analysis on disease cases can be found in our GitHub repository). We speculate that this might be due to a higher heterogeneity in the appearance of the disease cases. This links to a major bottleneck in the current feature importance analysis workflow, where the method can only show important input features per query, instead of directly deriving high level information from the trained model weights. For instance, the analysis cannot tell us directly from the trained model weights that the wall motion abnormality at the end systolic phase is the most critical sign for a disease. Human interpretation is still required to obtain insights from the trained model by going through the highlighted regions in multiple input queries. Another limitation of the currently available feature importance analysis methods is that the DeepLIFT analysis result is often quite noisy (22). This is especially true in our case, where the input ultrasound is already noisy. The noise would often hinder the further interpretation process. Another limitation of our study is the lack of patient-relevant information. Due to data protection regulations, all ultrasound data used in this study were anonymized and stripped of identifying metadata. Therefore, we were not able to maintain subject-level independence for the training-validation-testing splits, and quantitative research on the impact of age, sex, and medical history was not possible. If the metadata would become available in the future, the current analyses could be extended to investigate the possibility of integration of non-imaging data in the model such as ECG and other relevant clinical information or medical history. Finally, currently there are no other public datasets with diagnostic labels available as in our dataset. If, in the future, an independent validation dataset becomes available, we would be able to further verify the generalization ability of our models.

## Conclusions

In conclusion, this pilot study shows the feasibility of a 3D CNN approach in the detection of impaired LV function and AV regurgitation based on A4C-view ultrasound cineloops, which paves the way for an automated CVD diagnosis that can be made more accessible. Moreover, it demonstrated that deep learning methods can learn from large training data to detect diseases different from the predefined conventional way, and potentially discover diagnostic image features not previously paid attention to by humans.

# Figures and Tables

Table 1. Characteristics of the dataset.

Per diagnosis, the total number of ultrasounds extracted (# total) and ultrasounds per training, validation, and testing dataset are presented.

| Diagnosis | #Total | #Training | #Validation | #Test |
|---|---|---|---|---|
| Normal | 1888 | 1322 | 189 | 377 |
| Mildly impaired LV function | 509 | 356 | 51 | 102 |
| Severely impaired LV function | 419 | 293 | 42 | 84 |
| Mild AV regurgitation | 285 | 200 | 28 | 57 |
| Substantial AV regurgitation | 453 | 317 | 45 | 91 |

AV = aortic valve; LV = left ventricular.

Figure 1. Study overview. Two R(2+1)D models were trained to detect impaired LV function and AV regurgitation. Subsequently, tSNE was used to visualize the embedding of the extracted feature vectors, and DeepLIFT was used to identify important image features associated with the diagnostic tasks. (A4C = apical four chamber; AV = aortic valve; LV = left ventricular).

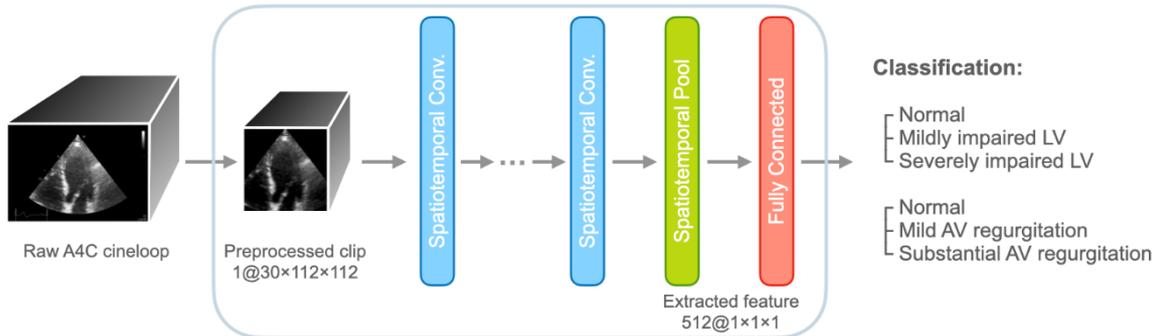

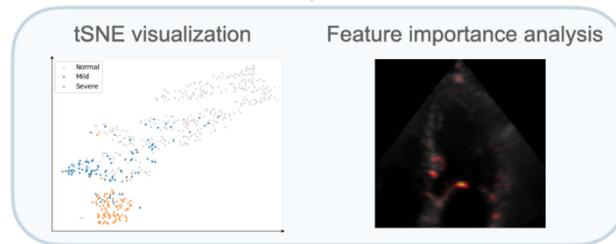

Figure 2. Predictive performance of the impaired LV function detection model (A, B) and the AV regurgitation detection model (C, D). (A) and (C) are the normalized confusion matrices for each classification task. (B) and (D) list the detailed recall (sensitivity), precision, and F1-score of each class. The performance demonstrates the feasibility of detecting the diseases using A4C cineloops.

(A)

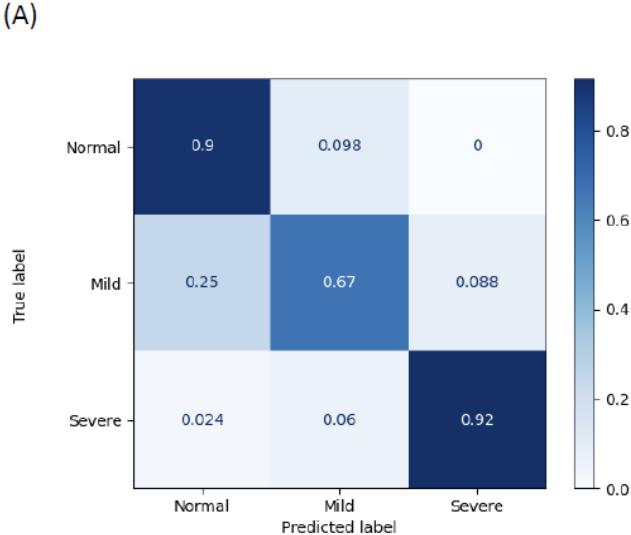

(B)

|  | Recall | Precision | F1-score |
|---|---|---|---|
| Normal | 0.9 | 0.93 | 0.91 |
| Mild | 0.67 | 0.62 | 0.64 |
| Severe | 0.92 | 0.9 | 0.91 |

(C)

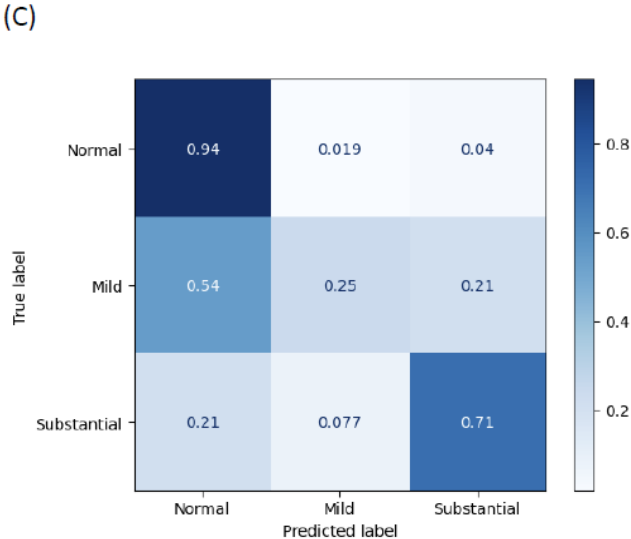

(D)

|  | Recall | Precision | F1-score |
|---|---|---|---|
| Normal | 0.94 | 0.88 | 0.91 |
| Mild | 0.25 | 0.5 | 0.33 |
| Substantial | 0.71 | 0.71 | 0.71 |

Figure 3. tSNE visualization for impaired LV function (A, B) and AV regurgitation (C, D). (A) and (C) are the visualizations of input ultrasound, in which the input video was directly reduced into a two-dimensional embedding by tSNE and visualized. (B) and (D) are the visualizations of the model-extracted feature vector, in which the 512-dimensional feature extracted by the model was reduced into a two-dimensional embedding by tSNE and visualized. Nearby dots in the plots imply similar data in the original high dimensional space. It can be observed in (B) and (D) that samples of the same diagnosis are clustered closer to each other, as compared to (A) and (C). This indicates that the models have filtered out irrelevant information and extracted diagnosis-relevant features. Especially, it can be seen in (B) that the model might have obtained information regarding the level of severity.

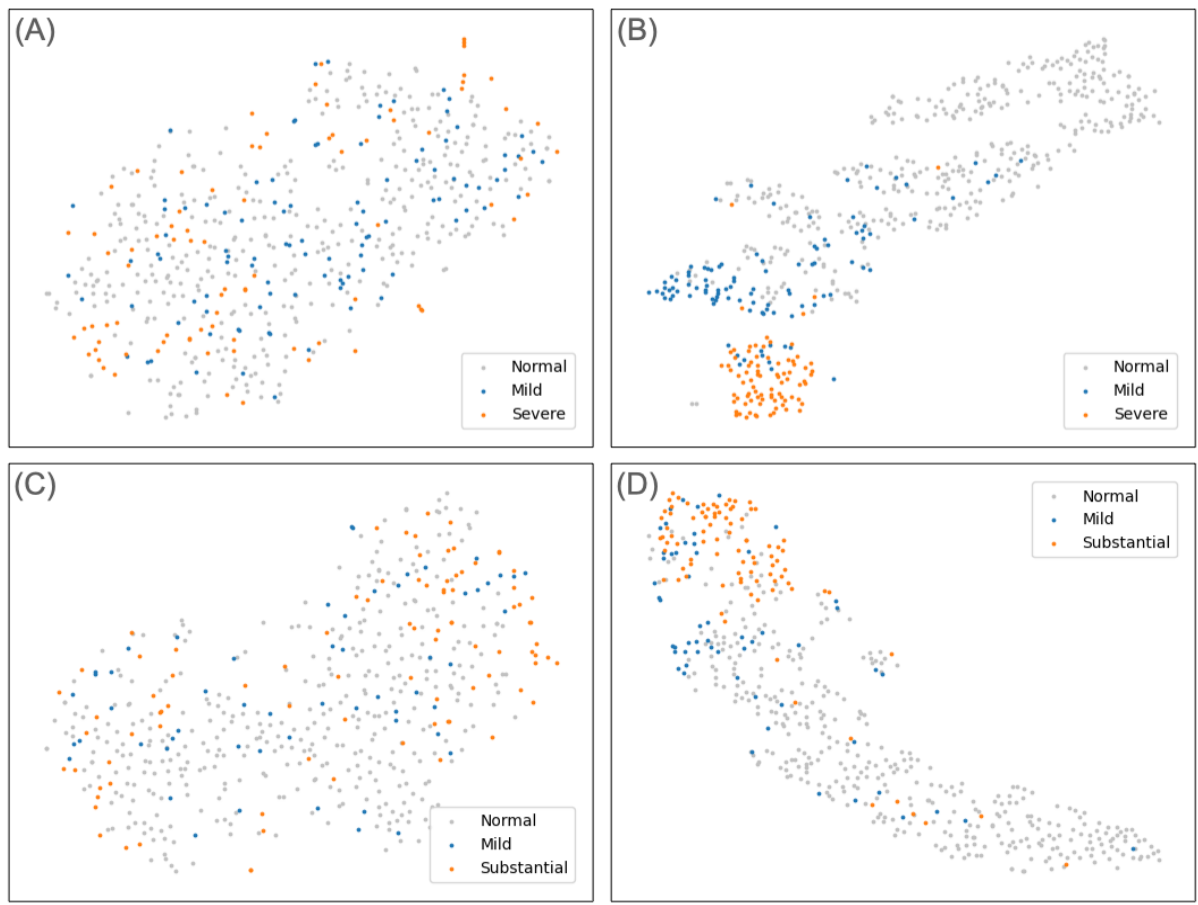

Figure 4. Feature importance analysis with DeepLIFT.

The highlighted regions are image features considered important by the respective models in order to distinguish normal cases from the disease cases. Panel (A) shows the analysis of the impaired LV function detection model. The LV myocardium at the basal level was highlighted from the early systolic to the early diastolic phase. The mitral valve was highlighted as well during the early diastolic phase. Panel (B) shows the analysis of the AV regurgitation detection model. The tip of the mitral valve anterior leaflet was highlighted particularly at a short time centering around the moment of valve opening. This indicates that the model focuses not only on a certain anatomical structure but also on a certain temporal phase within the cardiac cycle. This representative case shown here is the one with the highest average probability of being the normal class, as predicted by the two models, namely, a confident case. More examples in video format are available in our GitHub repository.

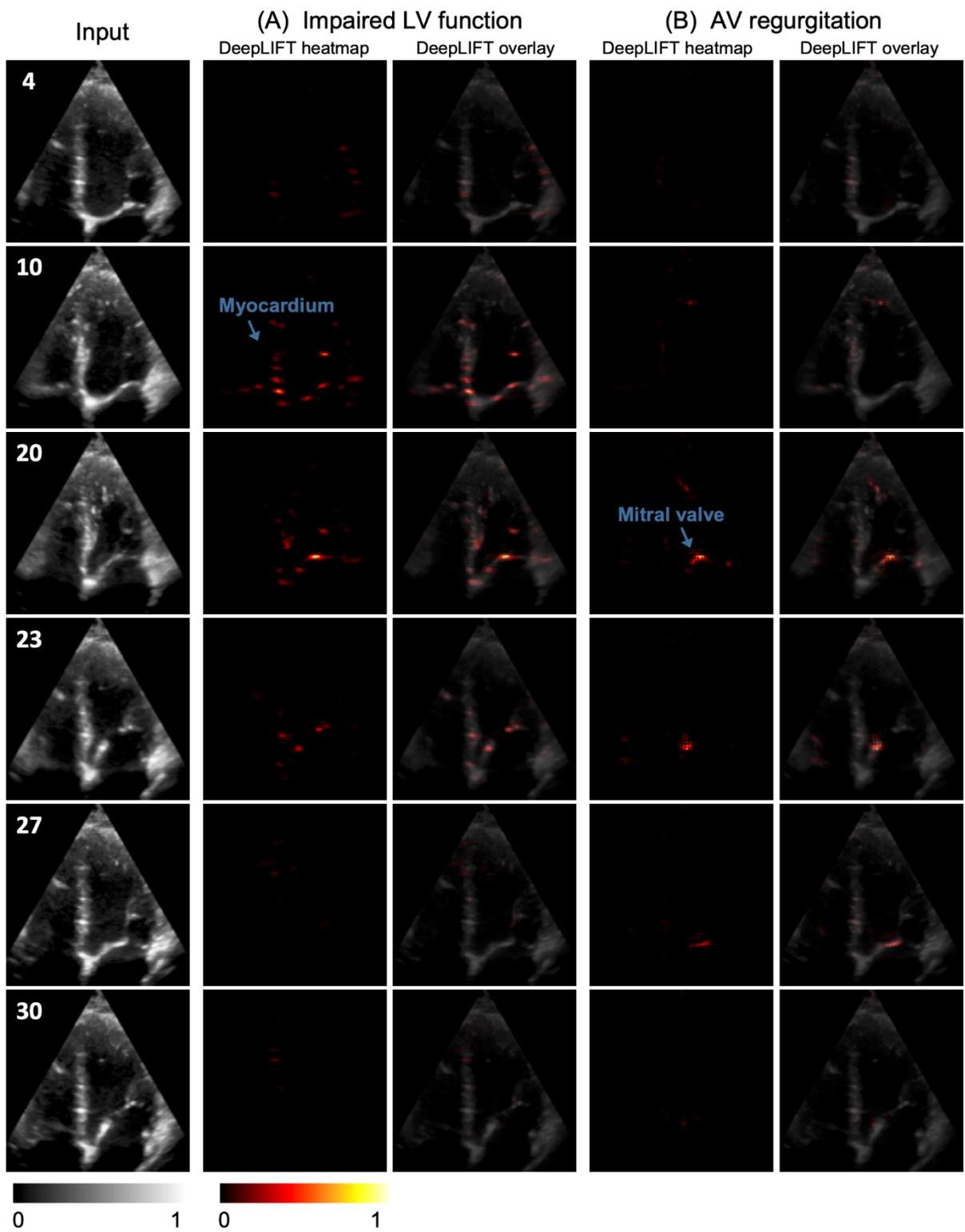